\documentclass[letterpaper,aps,prd,superscriptaddress,tightenlines,preprintnumbers,nofootinbib]{revtex4}
\usepackage{epsfig}
\usepackage{color}

\newcommand{\postscript}[2]{\setlength{\epsfxsize}{#2\hsize}
   \centerline{\epsfbox{#1}}}

\usepackage[usenames,dvipsnames]{xcolor}
\definecolor{rossoCP3}{cmyk}{0,.88,.77,.40}
\begin{document}

\title{\color{rossoCP3}{ Roadmap for Ultra-High Energy Cosmic Ray Physics and Astronomy} \\\color{rossoCP3}{ (whitepaper for Snowmass 2013)}}

\author{Luis A.~Anchordoqui}
\address{Department of Physics, University of Wisconsin-Milwaukee, Milwaukee, WI 53201, USA}
\author{Glennys R.~Farrar}
\address{Center for Cosmology and
  Particle Physics \& Department of Physics, New York University, New
  York, NY 10003, USA}
\author{John F. Krizmanic}
\address{NASA/Goddard Space Flight Center, Greenbelt, Maryland, 20771 USA}
\address{CRESST/Universities Space Research Association}
\author{Jim Matthews}
\address{Department of Physics and Astronomy, Louisiana State University, Baton Rouge, LA 70803, USA}
\author{John W. Mitchell}
\address{NASA/Goddard Space Flight Center, Greenbelt, Maryland, 20771 USA}
\author{Dave \nolinebreak  Nitz}
\address{Department of Physics, Michigan Technological University, Houghton, MI, USA}
\author{Angela \nolinebreak V. \nolinebreak  Olinto}
\address{Department of Astronomy and Astrophysics, Enrico Fermi
Institute, University of Chicago, Chicago, Il 60637, USA}
\address{Kavli Institute for Cosmological Physics,  University of Chicago, Chicago, Il 60637, USA}
\author{Thomas C. Paul}
\address{Department of Physics, University of Wisconsin-Milwaukee, Milwaukee, WI 53201, USA}
\address{Department of Physics, Northeastern University, Boston, MA 02115, USA}
\author{Pierre Sokolsky}
\address{Physics Department University of Utah Salt Lake City, Utah 84112 USA}
\author{Gordon B. Thomson}
\address{High Energy Astrophysics Institute \& Department of Physics and Astronomy, University of Utah, Salt Lake City, Utah, USA}
\author{Thomas J. Weiler}
\address{Department of Physics and Astronomy,
Vanderbilt University, Nashville TN 37235, USA
}

\maketitle

The origin and nature of the highest energy particles ever observed are
fundamental questions whose answers appear to be within our reach in the coming
decade.  The history of cosmic ray studies has witnessed many discoveries
central to the progress of high-energy physics, from the watershed
identification of new elementary particles in the early days to the confirmation
of long-suspected neutrino oscillations, to measuring cross-sections and
accessing particle interactions far above accelerator energies.  A major recent
achievement is establishing the suppression of the spectrum at the highest energies;
this may be the long-sought ``GZK'' cut-off predicted by Greisen, Zatzepin
and Kuzmin in 1966~\cite{Greisen:1966jv}, discussed in greater detail below.
The GZK suppression is a remarkable example of the profound links between
different regimes of physics, connecting as it does the behavior of the rarest,
highest-energy particles in the Cosmos to the existence of Nature's most
abundant particles -- the low energy photons in the relic microwave radiation of
the Big Bang -- while simultaneously demanding the validity of Special
Relativity over a mind-boggling range of scales.

Ultra-high energy cosmic rays (UHECRs), now commonly taken to be CRs with energies $> 6 \times
10^{19}$~eV, were first reported just over 50 years ago by John
Linsley~\cite{Linsley:1963km}.  These are the only particles with
energies exceeding those available at terrestrial accelerators. The
Large Hadron Collider (LHC) will reach an equivalent fixed-target
energy of $10^{17}$ eV, whereas UHECRs have been observed with
energies in excess of $10^{20}$~eV. With UHECRs one can conduct
particle physics measurements up to two orders of magnitude higher in
the lab frame, or one order of magnitude higher in the center-of-mass
frame, than the LHC energy reach. As discussed in more detail below,
the properties of UHECR air showers appear to be inconsistent with
models which are tuned to accelerator measurements; one possible
explanation is that new physics intervenes at energies beyond the LHC
reach. UHECR  experiments are the only way to access this energy range and make detailed measurements of air showers in order to address this question.  It is worth noting that cosmic ray experiments have already yielded particle
physics results at energies far exceeding those accessible to the LHC,
one of the latest being a measurement of the $p$-air cross-section at
$\sqrt{s} = 57~{\rm TeV}$~\cite{Collaboration:2012wt}, a result which
excludes some hadronic models extrapolations beyond LHC energies.


The two largest currently operating UHECR observatories are the Pierre Auger
Observatory in the Southern hemisphere, covering an area of 3000~km$^2$, and the
Telescope Array (TA) in the Northern hemisphere, covering about 700~km${^2}$.
Both observatories employ hybrid detection techniques, sampling cosmic ray air
shower particles as they arrive at the Earth's surface and also detecting the
fluorescence light produced when UHECR air showers excite atmospheric nitrogen,
for the $\sim 10$\% of events arriving on dark, moonless nights.  Both Auger and
TA feature ``low energy'' extensions, which will provide an overlap with the LHC
energy regime, while also allowing measurements in the galactic-to-extragalactic transition region.

\begin{figure}[ht]
\postscript{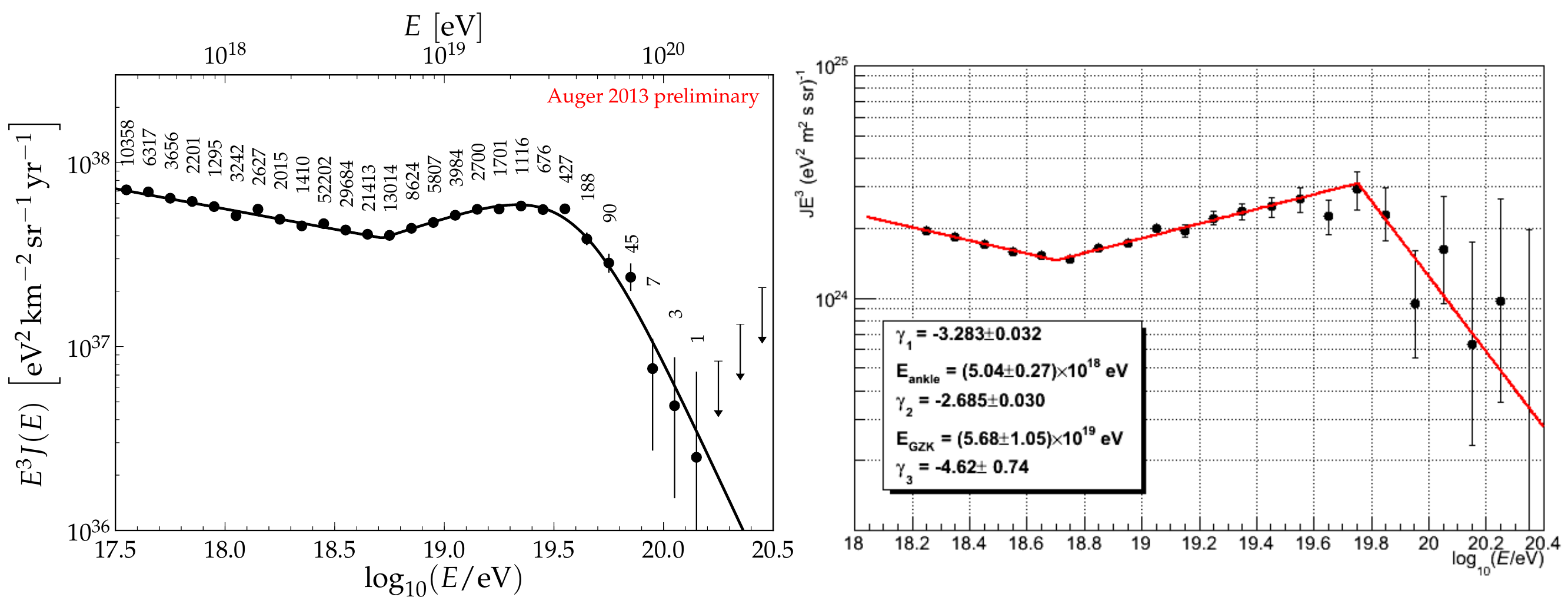}{0.9}
\caption{Energy spectra presented at ICRC 2013 by the Pierre Auger (left panel) and TA (right panel) collaborations. The Auger plot is labeled with the total number of events in each bin, with the last three arrows representing upper limits at 84\%~CL~\cite{ThePierreAuger:2013eja}. In the TA plot,
the legend gives the spectral indices and locations of energy breaks in a simple broken-power-law fit.}
\label{spectra}
\end{figure}

The most important result so far
from the present generation of observatories is the conclusive evidence that the
UHECR flux drops precipitously at high energy, as shown in Fig. \ref{spectra}.  The
discovery of a suppression at the end of the cosmic ray spectrum was first
reported by HiRes and Auger~\cite{Abbasi:2007sv,Abraham:2008ru} and later
confirmed by TA~\cite{AbuZayyad:2012ru}; by now the significance is well in
excess of 20$\sigma$ compared to a continuous power law extrapolation beyond the
ankle feature~\cite{Abraham:2010mj}.    This suppression is consistent with
the GZK prediction that interactions with cosmic background photons
will rapidly degrade UHECR energies~\cite{Greisen:1966jv}.   Intriguingly, however, there are also indications that
the source of the suppression may be more complex than originally anticipated.

Lower energy observations of the elongation rate (the rate of change with energy
of the mean
depth-of-shower-maximum, $X_{\rm max}$)~\cite{Bird:1993yi,AbuZayyad:2000ay,Tunka11,Knurenko:2011zz}, indicate
that the composition becomes lighter as energy increases toward $\sim 10^{18.3}$
eV from below, fueling a widespread supposition that extragalactic cosmic rays
are primarily protons.  However the Auger Observatory's high-quality,
high-statistics data sample exhibits a {\it decreasing} elongation rate as well
as a decreasing spread in $X_{\rm max}$ with increasing energy.
Interpreted with present shower simulations, this implies that the composition
is becoming gradually heavier beginning around $ 5 \times
10^{18}$~eV~\cite{Abraham:2010yv,Abreu:2013env}.  If true, this would have
important implications for the astrophysics of the sources.  A trend toward
heavier composition could reflect the endpoint of cosmic acceleration, with
heavier nuclei dominating the composition near the end of the spectrum -- which
coincidentally falls off near the expected GZK cutoff
region~\cite{Aloisio:2009sj}.  In this scenario, the suppression would
constitute an imprint of the accelerator characteristics rather than energy loss
in transit. It is also possible that a mixed or heavy composition is emitted
from the sources, and photodisintegration of nuclei and other GZK energy losses
suppress the flux~\cite{Allard:2011aa}.

An alternative possibility for the origin of the break in the
elongation rate could be even more interesting: this feature might
arise from some change in the particle interactions at UHE not
captured by event generators tuned to LHC and other accelerator data.
Adding weight to this possibility are the Auger measurements, using
three independent methods, showing that existing hadronic interaction
models do not simultaneously fit all shower observables.  For example,
the actual hadronic muon content of UHE air showers measured in hybrid
events is a factor $1.3-1.6$ larger than predicted by models tuned to
LHC data~\cite{AugerMuonICRC13}, even allowing for a mixed
composition.  Thus a critical step required to fully understand
$X_{\rm max}$ observations is to identify and correct the deficiencies
in current shower models.  This is a strong motivation for upgrading
present-generation detectors to enable full understanding of the
hadronic interactions involved in air shower development.
Fortunately, the information which will be accessible in shower
observations -- including the correlation between $X_{\rm max}$ and
the ground signal in individual hybrid events~\cite{Allen:2013hfa},
the comparison between $X_{\rm max}$ and $X_{\rm max}^{\mu}$ (the
atmospheric depth where muon production is maximum), the dependence of
ground signal on zenith angle, and other detailed shower observations
-- is so rich and multifaceted that it will enable composition and
particle physics to be disentangled~\cite{Allen:2013hfa}.

An additional intriguing twist in the present observational situation is that
the HiRes and TA results are consistent with a proton dominated flux everywhere
above the ankle~\cite{Abbasi:2009nf, Tsunesada:2011mp}, although with present
statistics the TA and Auger elongation rates agree within
errors~\cite{TA-AugerER-ICRC13}.  Since the sources seen by the HiRes and TA in the
Northern hemisphere may not the same sources as seen by the Auger Observatory in the South, the
composition need not be the same.  When TA statistics are sufficient to clearly
determine whether the elongation rate observed in the North is the same
as recorded by the Auger observatory  in the South, it will be of great consequence for
astrophysics even without knowing exactly how to translate from elongation rate
to composition.  If the composition (elongation rate) in North and South are not
the same, it will mean {\it i)} that there are at least two source types, one
accelerating primarily protons and another accelerating a mixed composition, and
{\it ii)} that in at least one hemisphere, the UHECRs are produced mainly by one
or a small number of sources.

Another major result of the present generation of observatories is the search for
anisotropy in the distribution of arrival directions.  Around $10^{18}$ eV, Auger has provided a strong
upper limit on the dipole anisotropy~\cite{Abreu:2012lva,Auger:2012an} which is
almost sufficient to rule out a Galactic origin assuming these cosmic rays are indeed
predominantly protons and making reasonable assumptions about the Galactic magnetic
field (GMF).  When the TA and Auger data are combined, the limit will be even stronger
or a signal will be found~\cite{delignyAugerICRC13}.

As the energy increases, evidence for anisotropy mounts. Auger has
reported a notable correlation of cosmic ray arrival directions with
nearby galaxies of the Veron-Cetty and Veron catalog of Active
Galactic Nuclei (AGN)~\cite{Cronin:2007zz}. With more data
accumulated, the central value of the correlation fraction has
decreased but the significance has remained at the 3-sigma
level~\cite{Abreu:2010ab,AugerICRC-HL12}.  The HiRes experiment did
not observe such a correlation~\cite{Abbasi:2008md}, but the most
recent results from TA ~\cite{AbuZayyad:2012hv,TinyakovTAanisoICRC13}
show a degree of correlation compatible with that seen by Auger in its
full data set, and with a similar pre-trial significance.
Furthermore, TA finds a significant correlation between the highest
energy events' pointing directions and the local large-scale structure
of the universe~\cite{TA-AugerAniso-ICRC13}.

While indications of anisotropy are becoming stronger, a completely clear
picture is thus far elusive, especially regarding the identity of the sources
themselves.  Perhaps a clear picture should not be expected, given the
possibility of multiple types of sources and the fact that the composition in
the South could be mixed or become heavy at the highest energies, while  the
flux could be more proton-dominated in the North.  Adding to the difficulty of
comparing correlation results of Northern and Southern hemisphere observatories
is the fact that the magnitude and directions of magnetic deflections and the
degree of multiple-imaging are expected to vary quite strongly across the
sky~\cite{fjr13}.  Fortunately astrophysics observations and theoretical effort
are rapidly improving GMF models~\cite{Jansson:2012pc,Jansson:2012rt}, so that the back-projection to
correct for deflections should become feasible, to some extent, on the
time-scale of the next generation of experiments.

Finally, we note the importance of limits which have already been placed on UHE photons and neutrinos.
Searches for UHE photons have rather dramatically changed our understanding of the early universe.  In particular,
topological defects and super heavy relics surviving to the present day
would decay to UHE photons and neutrinos. Existing UHECR experiments have
already provided powerful bounds on the UHE photon flux, ruling out many
existing models~\cite{photons}.  Future space missions will probe this
landscape even further, testing beyond-standard-model
physics models~\cite{Berezinsky:2011cp}.  Furthermore, though not designed specifically to detect
neutrinos, UHECR observatories have been able to achieve respectable neutrino
sensitivity~\cite{Abraham:2007rj} in an energy regime complementary to the
energy ranges in which dedicated neutrino experiments like IceCube and ANITA
achieve the best sensitivity.  \\

The next-generation UHECR observatories have three primary goals: \\
$\bullet$ {\it Increased statistics in both Northern and Southern
  hemispheres.}  A large increase in statistics is obviously
important to increase the
significance and resolution of all results.  In particular, it will
improve the chances of finding multiplets and resolving large scale
structure at higher energies, allow a more sensitive measurement of
the spectral suppression and potentially establish variations in the
spectrum in different regions of the sky. Furthermore, increased
statistics will aid in reducing systematic uncertainties (of all
sorts) for all measurements.
\\ $\bullet$ {\it
  Composition-tagging for each individual event.}  Probabilistic
composition-tagging for all events will address the question of how the
composition evolves with energy, thereby clarifying the nature of the
spectral cutoff and the acceleration mechanism(s).  It will also aid in source
identification by allowing events to be backtracked through the GMF, with reduced
ambiguity from their charge assignment, and allow correlation studies to be restricted
to proton-like events with smaller deflections.  \\ $\bullet$ {\it Detailed
  observations of UHECR showers to understand hadronic interactions in the UHE
  regime.}  It is essential to have reliable shower-development models and UHE
cross sections to be able to infer composition from the shower properties.
UHECRs are also Nature's highest energy particle beam and thus present an opportunity to
search for new types of particle physics.

Resolving the fundamental questions of UHECR composition and origins, and
investigating particle physics above accelerator energies, will require both
enhanced experimental techniques implemented at the existing observatories,
as well as a significant increase in exposure to catch the exceedingly rare highest
energy events. Pursuing improved ground-based detection techniques and
pioneering space-based observation will offer complementary tools to piece
together answers to these important but challenging puzzles.

The Auger Collaboration is preparing an upgrade proposal, to be
implemented starting in 2015~\cite{ThePierreAuger:2013mga}.  The
surface detector electronics will be improved to have 3 times better
time resolution and to allow the signal to be measured accurately much nearer to the shower core.  The surface detector stations
will be equipped to provide superior muon-electromagnetic (EM)
separation, either by segmenting the tank liners into an upper and
lower portion (taking advantage of the different energy deposition
characteristics of the EM and muon components) or by adding direct
muon or EM detectors underneath, on, or beside the existing tanks.  Not
only are these upgrades essential to clarify the composition puzzle
and elucidate the nature of the suppression, they will also allow us
to test the validity of existing hadronic interaction models and
possibly even reveal evidence of new physics at energies beyond the
reach of the LHC~\cite{Farrar:2013sfa}.   By enabling event-by-event muon reconstruction and composition tagging with the surface detector alone, the effective exposure for the precious composition-tagged events will increase by an order of magnitude, dramatically increasing the power to identify sources and test hadronic interactions.
The enhanced ability to reconstruct events, by exploiting the improved
understanding of shower physics, will be backwardly-applicable to
events observed with the original detector, increasing their value as
well. The Auger dataset will roughly triple in the next 10~years.

The TA experiment is proposing a project called TAx4 to increase the
size of its surface detector (SD) by a factor of 4 by adding 500
additional scintillation counters on a 2.08~km grid.  An additional
fluorescence detector (FD) site will be added overlooking the new
surface detectors.  The SD (FD) proposal will be made to the Japanese
(American) funding agencies. The size of TAx4 is matched to the TA
large-scale structure anisotropy signal, in that the larger detector
will be able to definitively observe (or rule out) the effect in three
years of running.  Also upcoming is a TA muon detector: TA is funded
to build, deploy, and operate a dedicated muon detector which will
augment our detailed current understanding of cosmic ray air
showers. Detectors of the TA Low Energy Extension (TALE) are currently
being deployed.  This will lower the low end of the TA energy range
from $10^{18}$~eV to $10^{16.5}$~eV, providing excellent coverage of
the LHC energy range, and accessing the astrophysics of the
galactic-to-extragalactic transition.

To maximize the utility of these existing and upgraded UHECR observatories, the Auger and TA teams have
established joint working groups to discuss experimental methods, pose
questions to one another on measurement techniques, compare data analyses and
modeling, and even share equipment.  This collaborative approach not only aids in
comparing results but also fosters a healthy environment of mutual evaluation
and constructive criticism of one another's techniques~\cite{UHECR-WGs}.

Moving beyond existing technologies, it is inspiring to note that some 5 million UHECRs above about $5.5 \times
10^{19}$~eV strike the Earth's atmosphere each year, from which we currently
collect only about 50 or so with present observatories. In this sense, there
exists some 5 orders of magnitude room for improvement! It may well be that
the best hope to make inroads in this area is to take the search for UHECR
sources into space, realizing a suggestion put forward by John Linsley in the
early 1980's~\cite{Benson:1981xe}.

The project closest to realizing this objective is the JEM-EUSO
mission~\cite{Adams:2013vea}, which is planned for launch no earlier than 2017
aboard the Kibo module of the International Space Station~\cite{jones}.
The primary objective of this mission is to launch a new era of particle
astronomy and astroparticle physics and potentially make the first individual
UHECR source identifications.  This instrument will employ wide field of view
Fresnel optics and a highly sensitive focal surface, complemented by a real-time
atmospheric monitoring system~\cite{kajino}.  A Global Light System comprising a
ground-based global network of calibrated light sources, operated remotely and
cross-checked by several aircraft flights per year, will validate and monitor
key parameters of the JEM-EUSO instrument over its planned
mission~\cite{Adams:2012hr}.  Operating in nadir (down-looking) mode,
JEM-EUSO mission will achieve, at $10^{20}$~eV, an annual exposure of $5.6 \times 10^{4}~{\rm km}^2 \,
{\rm sr \, yr}$ after factoring in the duty cycle, cloud coverage, and light
pollution estimates (assuming losses from reconstruction efficiency are minimal, as
anticipated).  The observatory will also be able to operate in tilt mode, increasing
the exposure and energy threshold: a 30$^{\circ}$ tilt will result in an annual
exposure of about $10^{5}~{\rm km}^2 \, {\rm sr \, yr}$~\cite{kenji-icrc}.  As
the ISS orbit is inclined at $51.6^\circ$, JEM-EUSO will enjoy uniform exposure
in both the Northern and Southern hemispheres, better facilitating searches for
full-sky anisotropy~\cite{tom-icrc} and other signs of the cosmic variance such
as ensemble fluctuations~\cite{Ahlers:2012az}. The instrument will also have the
capacity to distinguish extreme energy photons and neutrinos from baryonic
showers~\cite{nu-gamma}, though the mission will not add much to the debate over
baryonic mass composition discussed above. 

In addition to its physics potential, JEM-EUSO will serve as a
pathfinder for future space-based missions, establishing feasibility
and cost-effectiveness, uncovering challenges and opportunities, and
stimulating development of second-generation technology for more
ambitious projects. Even before we know the results from the upcoming
generation of UHECR observatories -- JEM-EUSO and upgraded Auger and
TA -- it seems clear that still larger aperture observatories with
much better energy and $X_{\rm max}$ resolution will be called for, in
order to measure the spectra and composition distribution of
individual sources.  The individual source spectrum is a key
diagnostic of the acceleration mechanism, being peaked from a bursting
source~\cite{Waxman:1996zn} and falling from a continuous one.
Furthermore, as the exciting IceCube cosmic neutrino observations have
recently shown~\cite{Aartsen:2013bka, Whitehorn}, we are entering the
era of neutrino astronomy, so the time is ripe for redoubled efforts
on this front.  Observation of cosmogenic neutrinos would make the
case for the reality of the GZK effect. Further, neutrino observations
can provide clues about the location of the galactic to extra-galactic
UHECR transition~\cite{Anchordoqui:2013qsi}.  Finally, if neutrinos
with energies above $10^{21}$~eV exist, JEM-EUSO and other satellite
instruments will have the potential to detect them.  Such an
observation would have dramatic implications, as the famous Hillas
acceleration constraints~\cite{Hillas:1985is} seem to exclude known
astrophysical objects from endowing neutrinos with such energies.

Strategies
and new devices to greatly enlarge terrestrial observatories are under
discussion, e.g., at the International Symposium on Future Directions
in UHECR Physics, CERN, Feb. 2012, and planning for space-based
observatories are underway. One such next-generation proposal is the OWL (Orbiting Wide-angle
Light-collectors) mission~\cite{Stecker:2004wt}, a pair of co-orbiting
satellites with f/1 Schmidt telescopes using deployable optics and
inflatable light shields, launched as a dual manifest on a single
rocket. Stereo event reconstruction, sub-pixel-crossing event timing,
and sophisticated atmospheric monitoring systems are expected to give
measurements nearly independent of track inclination and tolerant of
atmospheric conditions. In 1000~km low inclination orbits, the OWL
annual exposure is $2 \times 10^{5}~{\rm km}^2 \, {\rm sr \, yr}$
above $\sim 6 \times10^{19}$ eV. Initially, the spacecraft would fly
close together to detect Cherenkov light from upward-going neutrino
showers \cite{Krizmanic:2011} and then separate. A monocular mode
could double the detection aperture with the same energy threshold and
tilting the satellites could increase the aperture at higher
energies. Part of the mission at 600 km would lower the detection
threshold.

A study has also begun on GreatOWL~\cite{Krizmanic:2013} which will
employ inflatable optics $\sim 36$ times the OWL area together with
solid-state sensors at the focal surface, and more modest launch
requirements.  GreatOWL's energy threshold of $\sim 10^{18}$~eV, will
enable GZK-cosmogenic neutrino measurements and a significant
improvement $X_{\rm max}$ resolution, allowing composition
determination above $10^{19}$~eV~\cite{Krizmanic:2013b}. Inflatables
could allow JEM-EUSO or OWL class telescopes to use yet smaller
rockets for lower cost mission opportunities. Larger numbers of these,
possibly launched as multiple manifests, could form a
``constellation'' of UHECR telescopes.
Other space-based observatories have been proposed as possible concepts,
including quite recently the SWORD~\cite{Romero-Wolf:2013etm} concept, which
would employ satellite-born radio detection of air shower radio emissions
reflected off the Earth, possibly leading to observation of $\sim$100 events
above $~10^{20}$~eV per year in a cost-effective mission.

Some of the technologies needed to enable the next generation of UHECR
observatories are already in development. For example, there
are very active efforts underway to develop solid-state detectors for
future UHECR telescopes, for the Cherenkov Telescope Array and for direct
particle detectors~\cite{workshop}. The current focus is on
(SiPM)~\cite{Renker:2006ay}  optimized for UV with quantum efficiencies $\sim 60\%$
at 330~nm, over twice that of PMTs.

We also mention the interdisciplinary science derived from the
simultaneous function of these detectors as Earth observatories.  The
interdisciplinary science program at the Pierre Auger Observatory is
quite extensive~\cite{Lawrence}.  Through serendipity, Auger has
turned out to be the world's best detector for measuring atmospheric
transient luminous events known as Elves that are created above some
thunderstorms. Elves are part of the planet's electrical system. Their
detailed measurement by Auger provides a probe of the ionosphere.  In
the near future, the space-based TUS experiment~\cite{TUS} will
perform measurements of these atmospheric phenomena in the near-UV
while also serving as proof of principle for detection of UHECR with
enrgy $> 10^{20}~{\rm eV}$. Auger has also detected a major
earthquake, measured Forbush decreases associated with solar activity,
and developed many new techniques for measuring atmospheric
properties.


In conclusion -- thanks to a prodigious experimental effort -- the origin and
nature of the highest energy particles in the Universe are beginning to be
revealed.  Nonetheless, 50 years after their discovery much remains a mystery.
The way forward is clear and practical.  Upgrades to the Auger and TA
ground-based detectors, focusing especially on enhancing the capacity to infer
UHECR composition at the individual-shower level and improving our understanding
of UHE particle physics, will produce a major increase in science output at
modest cost.  Attaining the exposure necessary to pin down UHECR sources could
require taking the search to space, with the pioneering JEM-EUSO mission as our
current best opportunity. The high statistics achievable with JEM-EUSO will
complement the efforts of precision ground based experiments in revealing
the nature of the high energy flux suppression; for instance if the GZK effect
is indeed responsible for the observed suppression, the spectrum should display
a recovery if source injection energies exceed $5 \times 10^{20}~{\rm eV}$.
With the combined power of the space and ground-based approaches, a decade from
now we should be much closer to knowing what UHECRs are, where they come from,
and how they are produced; we may even have harnessed the study of UHECR
showers to explore particle physics at energies inaccessible to terrestrial
accelerators.\\


This work was supported by the US National Science Foundation (NSF),
the US Department of Energy (DoE) and the US National Aeronautics and
Space Administration (NASA).

\end{document}